\documentclass[preprint,showpacs,preprintnumbers,amsmath,amssymb]{revtex4}

\usepackage{graphicx}
\usepackage{dcolumn}
\usepackage{bm}
\begin{document}


\title{Hardness ratio evolutionary curves of gamma-ray bursts
expected by the curvature effect}

\author{Y.-P. Qin$^{1,2,3}$, C.-Y. Su$^{2,4}$ J. H. Fan$^{1}$, A. C. Gupta$^{2,5}$}
\altaffiliation{$^{1}$ Center for Astrophysics, Guangzhou University,
Guangzhou 510006, P. R. China \\
$^2$National Astronomical Observatories/Yunnan
Observatory, Chinese Academy of Sciences, Kunming 650011, P. R.
China \\
$^3$Physics Department, Guangxi University, Nanning 530004,
P. R. China \\
$^4$ Department of Physics, Guangdong Industry University,
Guangzhou 530004, P. R. China \\
$^5$ Tata Institute of Fundamental Research, Homi
Bhabha Road, Colaba, Mumbai 400 005, India} 
\email{ypqin@ynao.ac.cn, alok@ynao.ac.cn}

%

\date{\today}

\begin{abstract}
We have investigated the gamma-ray bursts (GRBs) pulses with
a fast rise and an exponential decay phase, assumed to
arise from relativistically expending fireballs, and found that
the curvature effect influences the evolutionary curve of the
corresponding hardness ratio (hereafter HRC). We find, due to the
curvature effect, the evolutionary curve of the pure hardness
ratio (when the background count is not included) would peak at
the very beginning of the curve, and then would undergo a
drop-to-rise-to-decay phase. In the case of the raw hardness ratio
(when the background count is included), the curvature effect
would give rise to several types of evolutionary curve, {\it
depending on the hardness of a burst.}
For a soft burst, an upside-down pulse of its raw HRC would be observed;
for a hard burst, its raw HRC shows a pulse-like profile with a sinkage in
its decaying phase; for a very hard burst, the raw HRC possesses a
pulse-like profile without a sinkage in its decaying phase. For a
pulse-like raw HRC as shown in the case of the hard and very hard
bursts, its peak would appear in advance of that of the
corresponding light curve, which was observed previously in some
GRBs. For illustration, we have studied here the HRC of GRB 920216, GRB
920830 and GRB 990816 in detail. The features of the raw HRC
expected in the hard burst are observed in these bursts. A fit to
the three bursts shows that the curvature effect alone could
indeed account for the predicted characteristics of HRCs. In
addition, we find that the observed hardness ratio tends to be
harder at the beginning of the pulses than what the curvature
effect could predict and be softer at the late time of the pulses.
We believe this is an evidence showing the existence of
intrinsic hard-to-soft radiation which might be due to the
acceleration-to-deceleration mode of shocks.
\end{abstract}

\maketitle

\section{Introduction}
Because of high output rate of observed radiation, gamma-ray bursts (GRBs)
are assumed to undergo a stage of fireballs which expand relativistically
[1,2]. Relativistic bulk motion of the gamma-ray-emitting plasma would
lead to some phenomena of GRBs [3]. To account for the observed spectra
of the bursts, the Doppler effect over the whole fireball surface would
play an important role [4-7].

There is a consensus that the prompt emission of GRBs might arise from
nonthermal synchrotron radiation [8,9], inverse Compton emission [10],
and others such as thermal, saturated Comptonization [11]. Among these
mechanisms, the nonthermal synchrotron is likely to be the dominant one
but this mechanism alone could not account for the spectra observed in
many GRBs. It is expected that in any case several mechanisms are at
work, the spectra would become complicated. Generally, a more flexible
one, the so-called Band function [12] which is purely an empirical
spectral form, is adopted to describe the observed spectra of most of
the bursts.

Some simple bursts with well-separated structure suggest that they
may consist of fundamental units of emission such as pulses, with
some of them being seen to possess a FRED form [13]. These FRED
pulses could be well represented by some empirical or
semi-empirical functions, and with these functions many
statistical properties of GRB pulses were revealed [14,15].
Phenomena of FRED pulses were interpreted as signatures of the
curvature effect [6,15-18]. {\it The effect arises from the
emission of the expanding surface of fireballs. Photons emitted
from different parts of the surface could reach a distant observer
at different times.} The delay of time is determined not only by
the difference of the corresponding emission time but also by the
variation of the distance of the emitting areas to the observer
when their angles to the line-of-sight are different. Meanwhile, in the
case that the angles to the line-of-sight of the emitting areas
are different, the Doppler shifting of the photons will differ
significantly. Combining all possible effects which are due to the
expending fireball surface comes to the full content of the
curvature effect (see a detailed analysis in [6]).

Light curves of GRBs in different energy channels differ
significantly in both the width and height of the curves [13]. In
the past few years, various attempts of interpretation of the
light curves have been made [14-17,19]. When fitting different
energy channel light curves, the different free parameters were
allowed due to the difference in appearances of the curves. Based
on these fits the temporal scale factors of a given pulse measured
at different energies were found to be related to the
corresponding energies by a power law [14,20-21]. It was suspected
that this behavior might result from a relative projected speed or
a relative beaming angle [21]. Due to the similar observed
features, the difference between different channel light curves of
GRBs might probably be due to the observed energy channel itself.
In other words, light curves of different energy channels might
arise from the same mechanism, with the only difference being the
energy range concerned. Indeed, as shown in [18], four channel
light curves of GRB 951019 could be well fitted by the same
mechanism (the same local pulse and the same rest frame spectrum)
when the curvature effect is taken into account. Other recent
investigations of the properties of FRED pulses of GRBs such as,
the width of GRB pulses as a power law function of energy, the
characteristics of the profile of FRED pulses in the decaying
phase, the spectral lag as a consequence of the curvature effect,
and the relationship between the power law index and the local
pulse width of FRED pulses were explored by [22-27]. Recently,
some authors attempted to explain the X-ray tails and flares
observed soon after the prompt emission of some GRBs with the
curvature effect [28].

The study of the spectral evolution of GRBs could be traced back to as early
as 1980s, where softening of the spectra with time as a general phenomenon,
the so-called hard-to-soft phenomenon was found [29]. Recently, quantitative
studies of the spectral behavior of GRB pulses were made by many authors
[17,30-33]. It was found that the hardness-intensity correlation obtained
from the statistical analysis was regarded as a result of the spectral
evolution, could be accounted for by the curvature effect [17]. However, as
a direct evidence of the spectral evolution, the evolutionary curve of the
spectral peak energy or the hardness ratio of GRB pulses should be intensely
explored and it would be necessary to check if the curvature effect
influences these curves.

In this paper, we study the evolution of the hardness ratio due to the fact
that the data of this quantity are available at time bins as small as $64$ms
in the BATSE catalogue. In this way, the details of the evolutionary curve
would be well illustrated. We first explore how the curvature effect
influences such theoretical curves for some typical bursts, and then choose
three FRED pulse sources to study this issue in order to check if individual
hardness ratio evolutionary curves could be accounted for by the curvature
effect.

\section{Theoretical prediction}

As suggested by many authors, the profile of GRB FRED pulses could be well
accounted for by the curvature effect [15-18]. We wonder what characteristics
of the HRC of FRED pulses could be expected when the curvature effect is at
work.

The formula employed is equation (21) presented in [18] which
could be written as
\begin{equation}
C(\tau )=C_0\frac{\int_{\widetilde{\tau }_{\theta ,\min }}^{\widetilde{\tau }%
_{\theta ,\max }}[\widetilde{I}(\tau _\theta )(1+\beta \tau _\theta
)^2(1-\tau +\tau _\theta )\int_{\nu _1}^{\nu _2}\frac{g_{0,\nu }(\nu
_{0,\theta })}\nu d\nu ]d\tau _\theta }{(1+\frac \beta {1-\beta }\tau )^2},
\end{equation}
with $\tau _{\min }\leq \tau \leq \tau _{\max }$, $\tau _{\min }\equiv
(1-\beta )\tau _{\theta ,\min }$, $\tau _{\max }\equiv 1+\tau _{\theta ,\max
}$, $\tau \equiv (t-D/c+R_c/c-t_c)/(R_c/c)$, and $\tau _\theta \equiv
(t_\theta -t_c)/(R_c/c)$, where $\theta $ is the angle to the line of sight,
$t$ is the observation time measured by the distant observer, $t_\theta $ is
the local time measured by the local observer located at the place
encountering the expanding fireball surface at the position denoted by angle
$\theta $ and the radius of the fireball $R=R(t_\theta )$, $t_c$ is the
initial local time, $R_c$ is the radius of the fireball measured at $%
t_\theta =t_c$, $D$ is the distance from the fireball to the observer, $%
\widetilde{I}(\tau _\theta )$ represents the development of the intensity
measured by the local observer, and $g_{0,\nu }(\nu _{0,\theta })$ describes
the rest frame radiation, and $\nu _{0,\theta }=(1-\beta +\beta \tau )\Gamma
\nu /(1+\beta \tau _\theta )$, $\widetilde{\tau }_{\theta ,\min }=\max
\{\tau _{\theta ,\min },(\tau -1+\cos \theta _{\max })/(1-\beta \cos \theta
_{\max })\}$ and $\widetilde{\tau }_{\theta ,\max }=\min \{\tau _{\theta
,\max },(\tau -1+\cos \theta _{\min })/(1-\beta \cos \theta _{\min })\},$
with $\tau _{\theta ,\min }$ and $\tau _{\theta ,\max }$ being the upper and
lower limits of $\tau _\theta $, which confine $\widetilde{I}(\tau _\theta )$%
, and $\theta _{\max }$ and $\theta _{\min }$ being the upper and lower
limits of $\theta $, which confine the emission area, respectively. With
this formula, count rates of both spherical fireballs (when $\theta _{\min
}=0$ and $\theta _{\max }=\pi /2$) and uniform jets (when $\theta _{\min }=0$
and $\theta _{\max }<\pi /2$) could be calculated as long as the local
emission intensity and the rest frame spectrum are assumed.

The spectral hardness ratio was previously defined as the ratio of the count
rate of a higher energy channel to that of a lower energy channel [29,34]. We
adopt this definition since it could be evaluated within a time interval of
$64ms$ and thus one can compare the theoretical curves with the observational
data of BATSE sources. The observed count subtracting the background count
within a $64ms$ interval for BATSE sources might become zero or negative due
to fluctuation. In this case, the hardness ratio would be ill-defined. To
avoid this difficulty, we simply study a raw hardness ratio which is defined
as the observed count of a higher energy band divided by that of a lower band,
where the background counts are not subtracted (in contrast to this, the ratio
of the observed count subtracting the background count of a higher energy band
to that of a lower band is called a pure hardness ratio). Following previous
studies, we define the raw hardness ratio as $HR_{3/2}\equiv C_3/C_2$,
where $C_3$ and $%
C_2$ are the expected or observed counts of the third and second BATSE
channels, respectively. The corresponding bi-channel light curves which is
defined as $C_{2+3}\equiv C_2+C_3$ will also be studied so that a direct
comparison between the hardness ratio curve and the light curve could be
made.

An expected value of count rates when taking into account the background
would be calculated with the light curve of (1) plus a background count rate
function (the expected background curve), and the expected value of the raw
hardness ratio would be calculated with this kind of count rate.

\begin{figure}
\includegraphics[width=5.0in,angle=0]{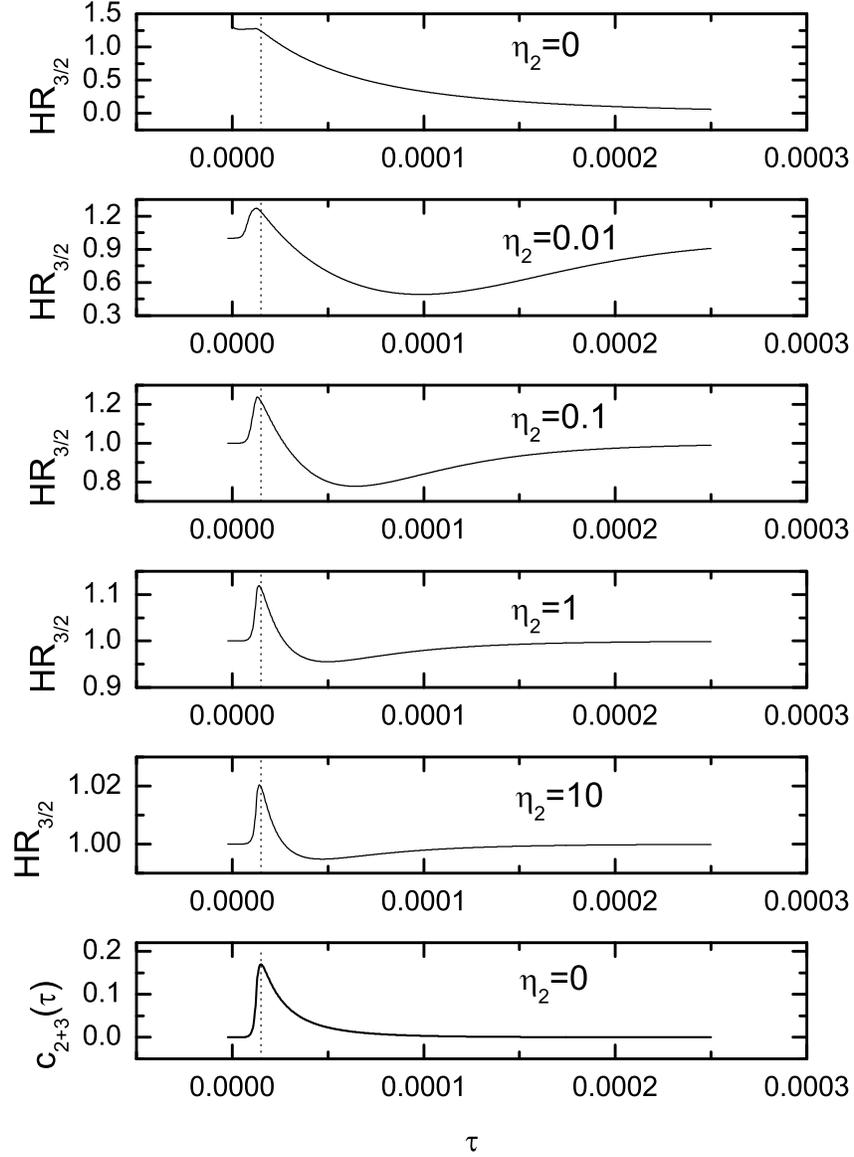}
\caption{\label{fig. 1}Plot of the hardness ratio $HR_{3/2}$ vs.
the relative observation time $\tau$ for various values of
$\eta_2$ (from top, the first to the fifth panels) in the case of
adopting $\Gamma=200$ and $\eta_3 = \eta_2$, where equations
(1)-(5) are employed and we take $\sigma _r=0.1$, $\sigma
_d=\sigma _r$, $\alpha _{0,C}=-0.6$, $\nu _{0,C} = 0.55 keV
h^{-1}$, and $C_0=1$. Plotted in the last panel is the
corresponding bi-channel light curve, where background counts are
not included. The dotted lines in all panels mark the time when
the peak count of the light curve presented in the last panel is
observed.}
\end{figure}

To reveal the main characteristics of the HRCs expected from the curvature
effect, we temporarily ignore possible effects arising from any of the
following factors: the real forms of the local pulse and the rest frame
spectrum, the development pattern of the rest frame spectrum, and the
variation mode of the Lorentz factor. We assume an exponential rise and
exponential decay local pulse
\begin{equation}
\widetilde{I}(\tau _\theta )=I_0\{
\begin{array}{c}
\exp (\frac{\tau _\theta -\tau _{\theta ,0}}{\sigma _r})\qquad \qquad \qquad
(\tau _{\theta ,\min }\leq \tau _\theta \leq \tau _{\theta ,0}) \\
\exp (-\frac{\tau _\theta -\tau _{\theta ,0}}{\sigma _d})\qquad \qquad
\qquad \qquad \qquad (\tau _{\theta ,0}<\tau _\theta )
\end{array}
,
\end{equation}
emitting with a rest frame Comptonized radiation form
\begin{equation}
g_{0,\nu }(\nu _{0,\theta })=\nu _{0,\theta }^{1+\alpha _{0,C}}\exp (-\nu
_{0,\theta }/\nu _{0,C}).
\end{equation}
We assign $\tau _{\theta ,0}=10\sigma _r+\tau _{\theta ,\min }$ so that the
interval between $\tau _{\theta ,0}$ and $\tau _{\theta ,\min }$ would be
large enough to make the rising part of the local pulse close to that of the
exponential pulse. Note that, $\tau _{\theta ,\min }$ could be freely chosen
so long as it satisfies the following constraint: $\tau _{\theta ,\min
}>-1/\beta $ [18]. Without any loss of generality we take $%
\tau _{\theta ,\min }=0$.

\begin{figure}[tbp]
\begin{center}
\includegraphics[width=5.0in,angle=0]{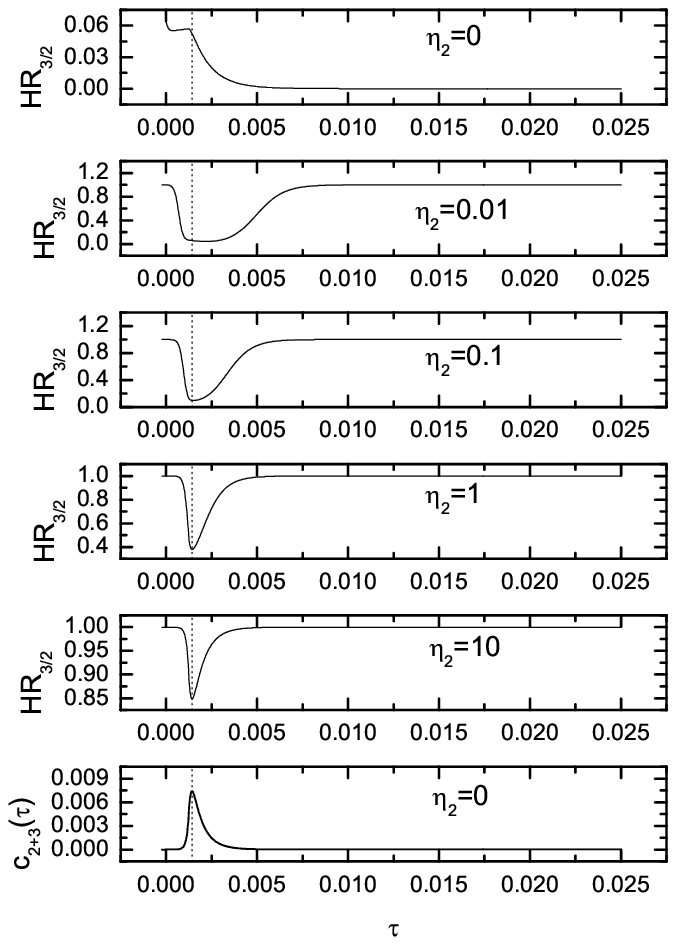}
\end{center}
\caption{Plot of the curves of Fig. 1 when calculating them by adopting $%
\Gamma=20$ instead of $\Gamma=200$.}
\label{Fig. 2}
\end{figure}

\begin{figure}[tbp]
\begin{center}
\includegraphics[width=5.0in,angle=0]{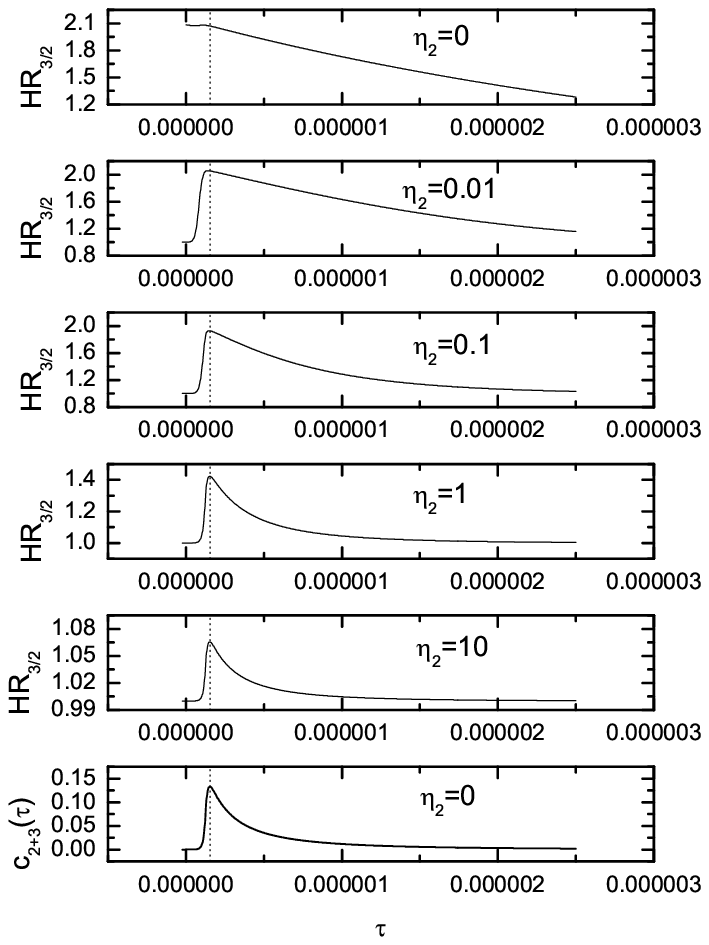}
\end{center}
\caption{Plot of the curves of Fig. 1 when calculating them by adopting $%
\Gamma=2000$ instead of $\Gamma=200$.}
\label{Fig. 3}
\end{figure}

The expected observed light curves in BATSE channels 2 and 3 are assumed to
be
\begin{equation}
C_{2,ob}(\tau )=C_2(\tau )+\eta _2\frac{C_{2+3,p}}2
\end{equation}
and
\begin{equation}
C_{3,ob}(\tau )=C_3(\tau )+\eta _3\frac{C_{2+3,p}}2
\end{equation}
where $C_2(\tau )$ and $C_3(\tau )$ are determined by (1) [note that $%
C_2(\tau )$ and $C_3(\tau )$ are different in the limits of the integral of
energy in equation (1)], $C_{2+3,p}$ is the peak of $C_{2+3}(\tau )$, and $%
\eta _2$ and $\eta _3$ are constants which denote the magnitude of
the background count of channels 2 and 3 relative to $C_{2+3,p}$
respectively. Here, for the sake of simplicity, we assume that the
background count does not evolve with time. Following [23], we
adopt $\alpha _{0,C}=-0.6$ and $\nu _{0,C }=0.55keVh^{-1}$ and
consider a typical hard and typical soft bursts with $\Gamma =200$
and $\Gamma =20$ respectively. In addition, a typical very hard
burst with $\Gamma =2000$ is also considered.

Displayed in Fig. 1 are the HRCs for the typical hard burst ($\Gamma =200$)
in the case of $\eta _3=\eta _2$ for various values of $\eta _2$. When
taking $\Gamma =20$ and $\Gamma =2000$ we get Figs. 2 and 3 for the typical
soft and typical very hard bursts respectively in the same case. For the
pure hardness ratio, the peaks of HRCs of the three typical bursts are $0.063
$ (for the typical soft burst), $1.3$ (for the typical hard burst) and $2.1$
(for the typical very hard burst). We find from these figures
that: a) for the signal alone (i.e., when the background count is not
included), the hardness ratio (here, the pure hardness ratio) would peak at
the very beginning of the light curve, and then would undergo a
drop-to-rise-to-decay phase, for the three bursts; b) for the typical soft
burst, an absorption-line-feature pulse (called an upside-down pulse) is
observed in its raw HRCs (i.e., when the background count is included), and
the larger the background count, the narrower the upside-down pulse; c) for
the typical hard burst, a pulse-like profile with a sinkage in its decaying
phase would be observed in its raw HRCs, and the larger the background
count, the shallower the sinkage; d) for the typical very hard burst, a
pulse-like profile without a sinkage in its decaying phase would be observed
in its raw HRCs, and the larger the background count, the narrower the
pulse; e) in the case of the two hard bursts, the peak of the raw HRCs would
appear in advance of that of the corresponding bi-channel light curve; f)
for the three typical bursts, the profile of the raw HRCs would depend
strongly on the ratio of the background count to the signal count.

To find out how background counts in the two channels are at work, we
calculate the HRCs of the typical hard burst in cases of $\eta_3 = 0.8
\eta_2 $ and $\eta_3 = 1.2 \eta_2$ as well. The results are presented in
Fig. 4. We find that the profile as well as the magnitude of the raw HRCs
would also depend on the ratio between the background counts in the two
corresponding channels.

\begin{figure}[tbp]
\begin{center}
\includegraphics[width=5in,angle=0]{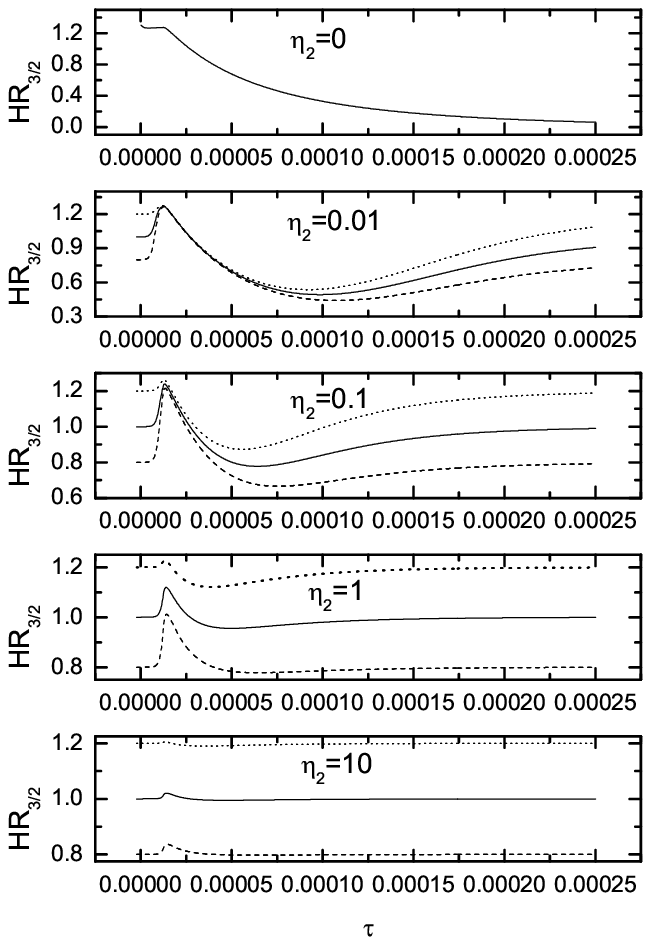}
\end{center}
\caption{Plot of the hardness ratio $HR_{3/2}$ vs. the relative observation
time $\tau$ for various values of $\eta_2$ in the case of adopting $\eta_3 =
0.8 \eta_2$ (dashed lines), $\eta_3 = \eta_2$ (solid lines), and $\eta_3 =
1.2 \eta_2$ (dotted lines), respectively. Other parameters are the same as
those adopted in Fig. 1.}
\label{Fig. 4}
\end{figure}

In Table 1, we have listed the peak time and width of the HRCs and light curves
of the typical hard burst in the case of adopting $\sigma _d=\sigma _r$ and
various sets of parameters. In calculating the width of the HRCs, we
consider only the portion of the curves where the values of the hardness
ratio are larger than that arising merely from the background counts (e.g.,
the flat section shown in the beginning of the HRCs in Figs. 1 and 4).
For the pure HRCs, since they are not a pulse-like curve, the width is not
available.

Adopting $\sigma _d=5\sigma _r$, we find the same characteristics observed
above, suggesting that the characteristics do not come from the form of
local pulses.

To investigate how the rest frame spectrum affects HRCs, we repeat
the above analysis by replacing the adopted Comptonized radiation
form with a Band function spectrum [12] with $\alpha
_{0}=-1$, $\beta _{0}=-2.25 $ and $\nu _{0,p} = 0.75 keV h^{-1}$.
We reached the same conclusions.

\begin{table}[tbp]
\caption{Peak time and width of the HRCs and light curves of the typical
hard burst in the case of adopting $\sigma _d=\sigma _r$ and various sets of
parameters}
\begin{center}
\tabcolsep0.21in
\begin{tabular}{llllll}
\hline\hline
curve & $\sigma_r$ & $\eta_3$ & $\eta_2$ & $\tau_{peak}$ & $FWHM_{\tau}$ \\
\hline
$C_{2+3}$ & $0.1$ & $0$ & $0$ & $1.49\times10^{-5}$ & $1.36\times10^{-5}$ \\
$HR_{3/2}$ & $0.1$ & $0$ & $0$ & $2.50\times10^{-8}$ & --- \\
$HR_{3/2}$ & $0.1$ & $0.8 \eta_2$ & $0.01$ & $1.27\times10^{-5}$ & $%
1.72\times10^{-5}$ \\
&  &  & $0.1$ & $1.34\times10^{-5}$ & $1.41\times10^{-5}$ \\
&  &  & $1$ & $1.42\times10^{-5}$ & $1.09\times10^{-5}$ \\
&  &  & $10$ & $1.44\times10^{-5}$ & $9.60\times10^{-6}$ \\
$HR_{3/2}$ & $0.1$ & $\eta_2$ & $0.01$ & $1.27\times10^{-5}$ & $%
1.21\times10^{-5}$ \\
&  &  & $0.1$ & $1.32\times10^{-5}$ & $9.60\times10^{-6}$ \\
&  &  & $1$ & $1.39\times10^{-5}$ & $8.08\times10^{-6}$ \\
&  &  & $10$ & $1.42\times10^{-5}$ & $7.32\times10^{-6}$ \\
$HR_{3/2}$ & $0.1$ & $1.2 \eta_2$ & $0.01$ & $1.24\times10^{-5}$ & $%
7.07\times10^{-6}$ \\
&  &  & $0.1$ & $1.27\times10^{-5}$ & $4.80\times10^{-6}$ \\
&  &  & $1$ & $1.32\times10^{-5}$ & $3.79\times10^{-6}$ \\
&  &  & $10$ & $1.34\times10^{-5}$ & $3.79\times10^{-6}$ \\ \hline
$C_{2+3}$ & $1$ & $0$ & $0$ & $1.44\times10^{-4}$ & $9.34\times10^{-5}$ \\
$HR_{3/2}$ & $1$ & $0$ & $0$ & $2.50\times10^{-7}$ & --- \\
$HR_{3/2}$ & $1$ & $0.8 \eta_2$ & $0.01$ & $1.27\times10^{-4}$ & $%
1.21\times10^{-4}$ \\
&  &  & $0.1$ & $1.32\times10^{-4}$ & $9.85\times10^{-5}$ \\
&  &  & $1$ & $1.39\times10^{-4}$ & $7.32\times10^{-5}$ \\
&  &  & $10$ & $1.42\times10^{-4}$ & $6.57\times10^{-5}$ \\
$HR_{3/2}$ & $1$ & $\eta_2$ & $0.01$ & $1.27\times10^{-4}$ & $%
9.09\times10^{-5}$ \\
&  &  & $0.1$ & $1.29\times10^{-4}$ & $6.82\times10^{-5}$ \\
&  &  & $1$ & $1.34\times10^{-4}$ & $5.56\times10^{-5}$ \\
&  &  & $10$ & $1.37\times10^{-4}$ & $5.05\times10^{-5}$ \\
$HR_{3/2}$ & $1$ & $1.2 \eta_2$ & $0.01$ & $1.24\times10^{-4}$ & $%
5.30\times10^{-5}$ \\
&  &  & $0.1$ & $1.27\times10^{-4}$ & $3.54\times10^{-5}$ \\
&  &  & $1$ & $1.29\times10^{-4}$ & $2.53\times10^{-5}$ \\
&  &  & $10$ & $1.32\times10^{-4}$ & $2.27\times10^{-5}$ \\ \hline\hline
\end{tabular}
\end{center}
\end{table}

\section{Observational characteristics of the hardness ratio evolutionary
curve shown in three FRED pulse bursts}

Since FRED pulses of GRBs are likely to suffer the curvature effect
[6,15-18,22-25], we suspect that the characteristics predicted above can be
checked by the observational data of the HRCs of such GRBs. Here, we
study the development of the hardness ratio for three bursts whose light
curves are that of FRED and then check if the features predicted above are
observed. The three bursts are chosen to investigate the issue due to the
following reasons: a) if a characteristic is common for most FRED pulses it
would likely be observed in fewer cases; b) if the curvature effect plays a
role in the hardness ratio evolutionary curve of a few bursts it would
probably play a role in other bursts of the same kind; c) as a primary
investigation, we concern in this paper only the characteristics of the
hardness ratio evolutionary curve revealed by individual pulses rather than
concern a statistical property of the bursts.

The bursts studied are GRB 920216 (\#1406), GRB 920830 (\#1883), and GRB
990816 (\#7711). Count rates of the bursts are available in the website of
BATSE [35,36],
where the counts within the $64ms$ bin for four energy
channels $[25,55]keV$, $[55,110]keV$, $[110,320]kev$, and $[>320]keV$ are
presented.

Presented in the BATSE website, one also finds the values of $T_{90}$, the
duration of a source, for the three bursts. We extend the signal data of the
sources from $T_{90}$ to $2T_{90}$, starting from $T_{90}/2$ previous to the
start time of the duration. Let this range be denoted by $t_{\min }\leq
t\leq t_{\max }$. Data beyond this range (that of $t<t_{\min }$ and $%
t>t_{\max }$), which are independent of the signal data, are employed to
find the fit to the count rate of the background.

The duration as well as the range of the $2T_{90}$ signal data of the three
bursts are as follows.

GRB 920216: $T_{90}=19.968s$, $t_{\min }=-9.28s$, $t_{\max }=30.656s$.

GRB 920830: $T_{90}=17.344s$, $t_{\min }=-8.288s$, $t_{\max }=26.4s$.

GRB 990816: $T_{90}=20.928s$, $t_{\min }=-10.016s$, $t_{\max }=31.84s$.

\begin{figure}[tbp]
\centering
\hspace*{1in}
\includegraphics[width=2.0in]{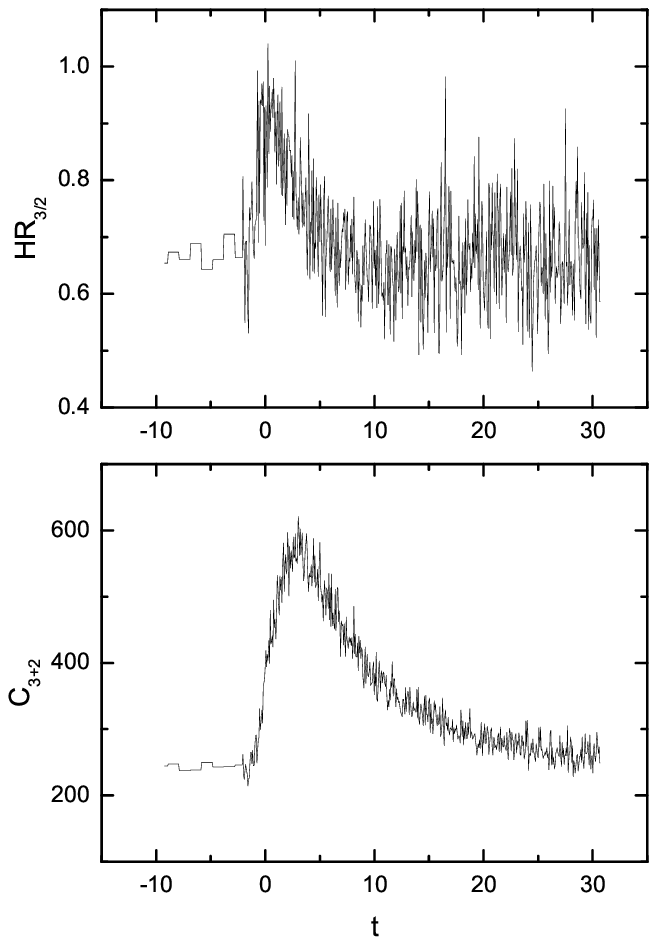}
\hspace*{1in}
\includegraphics[width=2.0in]{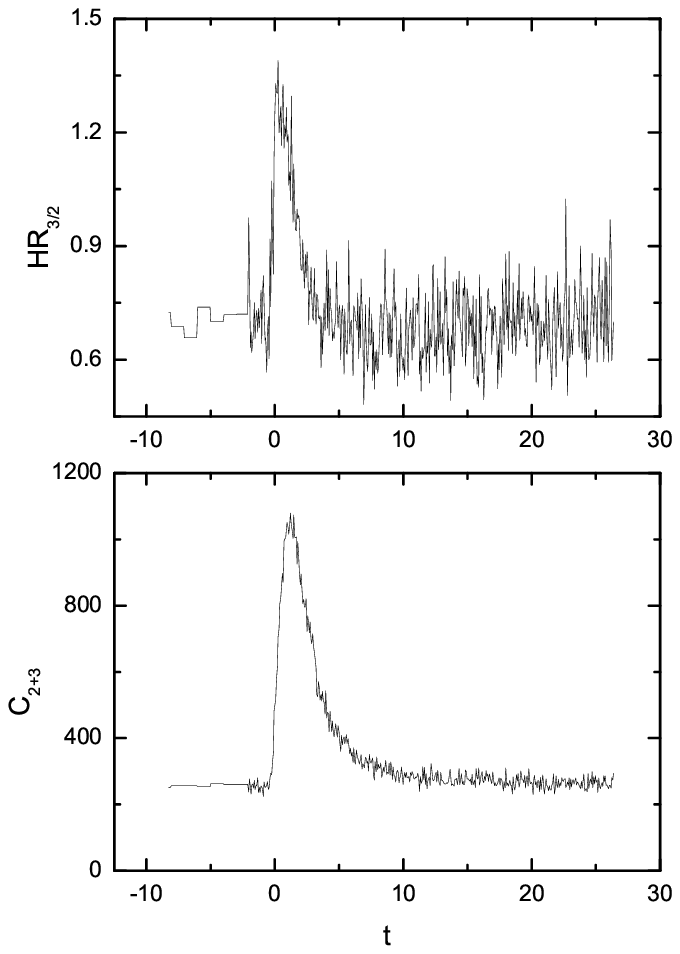}
\hspace*{1in}
\includegraphics[width=2.0in]{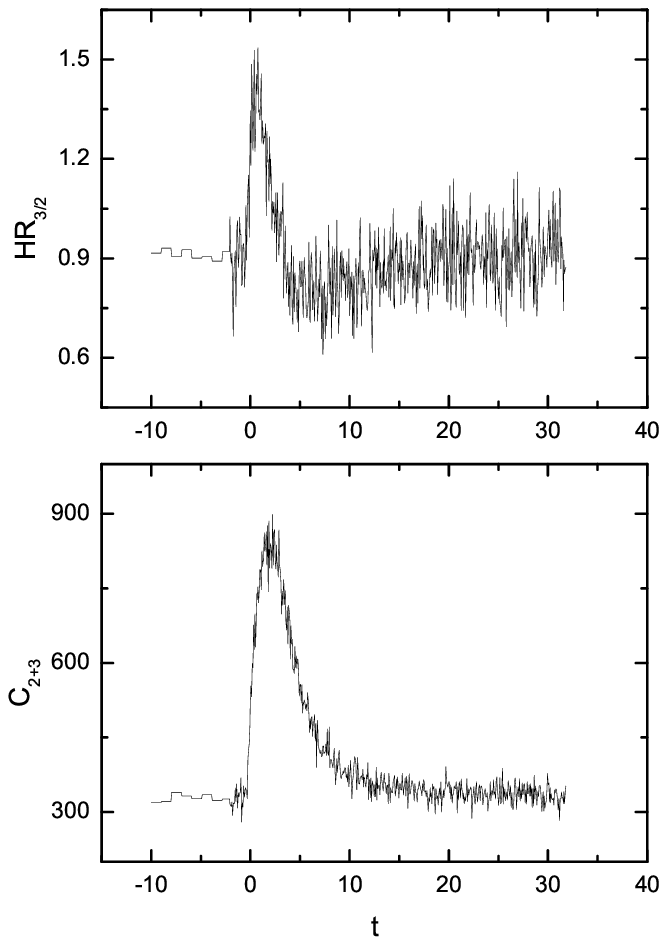}
\caption{Developments of the raw hardness ratios (upper panels) and the
bi-channel light curves which includes count rates of both the second and
third channels (lower panels) of GRB 920216 (top left),
GRB 920830 (top right), and GRB 990816 (bottom).}
\end{figure}

The developments of the raw hardness ratios $HR_{3/2}$
and the bi-channel light curves $C_{2+3}$ of the three bursts are shown
in Fig 5. Several
characteristics could be observed from the figure, with some of them being
reported previously [14,29,34,37]: a) the raw hardness ratio as a function
of time exhibits a pulse shape; b) the peak of the raw hardness ratio $HR_{3/2}$
appears in advance of the peak of the bi-channel light curve $C_{2+3}$; c)
the width of the HRC is narrower than that of the corresponding light curve;
d) there is a sinkage in the decay phase of the HRC. All these features are
in agreement with what predicted in the case of the typical hard burst (see
Fig. 1). Note that, the third character could be observed in the 4th and 5th
panels counting from the top in Fig. 1. It corresponds to the case of
relatively large background counts.

\section{Curvature effect as a key factor accounting for the hardness ratio
evolutionary curve}

As long as the characteristics observed in the three FRED pulse bursts are
in agreement with what predicted by the curvature effect, we try to make
a fit to the HRCs of these bursts based on the assumption that the curvature
effect is important. The formula employed is equation (1).

One could observe from Fig. 5 that the hardness ratio data are quite
randomly scattered. This is due to the much scatter of the count rate data.
There are several ways to ease this scattering. One is to smooth the count
rate data before the hardness ratio is calculated. Another is to deal with
the data associated with larger bins of time. With the former method one
could maintain the large number of data points but the information of signal
would be somewhat influenced, while with the latter method the information
of signal would not be affected, but the number of data points would be
reduced. As there are enough data points for the three bursts, we prefer the
latter approach.

The background data (that of $t<t_{\min }$ and $t>t_{\max }$) would be
fitted with a quadratic polynomial, and then this fitting curve would be
applied to the signal interval and would be taken as the expected background
count rate there (instead of finding it from websites, we perform the fit to
the background count ourselves since data of larger bins are dealt with in
this paper). An expected value of the count rate within the signal interval
when taking into account the background would be calculated with the light
curve of (1) plus the background count rate function (the background fitting
curve), and the expected value of the raw hardness ratio would be calculated
with this kind of count rate.

To check if the curvature effect could play an important role in producing
the HRC characteristics observed in the three bursts, we temporarily ignore
possible effects arising from the real forms of the local pulse, the rest
frame spectrum, the development pattern of the rest frame spectrum, or the
variation mode of the Lorentz factor. We therefore assume and employ the local
pulse (2) and rest frame spectrum (3) to perform the fit. Due to the same
reasons, some parameters adopted above are maintained, i.e., we assign $\tau
_{\theta ,0}=10\sigma _r+\tau _{\theta ,\min }$ and take $\tau _{\theta
,\min }=0$ (note that, the origin of time would be determined by $t_0$
presented below, and thus $\tau _{\theta ,\min }$ could be freely chosen).

As shown in [38], the opening angle of jets could be as
small as $\theta =0.03$. When applying equation (1), we take $\theta _{\max
}=0.03$ so that one could find if the HRCs observed above could arise from
uniform jets with such a small opening angle when the curvature effect is
taken into account. Since the profile of light curves is not significantly
affected by the Lorentz factor [18], we simply take $%
\Gamma =200$. To relate the observation time $t$ and the relative
time scale $\tau $, we assign $t=t_1\tau +t_0$. Since the
background counts are included in the definition of the raw
hardness ratio, the magnitude of counts, $C_0$, could not be
cancelled. Therefore, there would be seven free parameters
determined by fit, which are the rest frame peak energy $E_{0,C}$
and spectral index $\alpha _{0,C}$, the two widths of the
exponential rise and exponential decay local pulse $\sigma _r$ and
$\sigma _d$, the two time constants $t_1$ and $t_0$, and the count
magnitude $C_0$. {\it Note that the fitted parameters such as the
Lorentz factor are merely parameters of the model}. They are not
the real parameters of the bursts concerned. In particular,
adopting $ \Gamma =200$ as a parameter of the model does not imply
that the matter concerned must have such a large speed of motion.

We know that both the HRC and bi-channel light curve in Fig. 5 are
determined by the light curves of the second and third channels. If the
curvature effect is at work, it should affect not only the second and third
channel light curves but also the first and fourth channel light curves. We
thus regard relations arising from the four channel light curves calculated
with equation (1) as those associated with the curvature effect (here, as
stated above, other possible effects are ignored). Therefore, the observed
four channel light curves of the bursts will be fitted simultaneously with
equation (1) to find out if and how the observed HRC could be accounted for
by the curvature effect.

As shown in Fig. 5, the observed light curve data are quite randomly
scattered. We assume that fluctuation could be responsible to this
scattering. Based on this assumption, the statistics $\chi ^2$ associated
with a light curve could be defined as
\begin{equation}
\chi ^2\equiv \sum_{i=1}^n\frac{(C_{ob,i}-C_i)^2}{C_i},
\end{equation}
where $n$ is the total number of bins, $C_{ob,i}$ and $C_i$ are the observed
and expected values of the count respectively, within the $i$-th bin. The $%
\chi ^2$ of the light curve could then be calculated with equation (6). As
four channels are involved, we define a combined $\chi ^2$ to perform the
fit. Let the combined $\chi ^2$ be $\chi _{com}^2=\sum_{j=1}^4\chi _j^2$,
where $\chi _j^2$ is the $\chi ^2$ associated with the fit to the $j$-th
channel light curve. The numbers of bins for the four light curves are the
same for each source, and therefore the total number of data for $\chi
_{com}^2$ is $4n$ (note that data of different channels are independent).
The best fit would be obtained when the smallest value of $\chi _{com}^2$ is
reached. In this way, the four channel light curves could be ``
simultaneously'' fitted.

\textbf{GRB 920216:}

There are 7504 data points in total for GRB 920216. We divide this range of
data uniformly into 1876 bins with each bin containing four $64ms$ time
intervals. The quadratic polynomial fits to the background counts yield: $%
0.256C(t)=752-0.0812t-0.000500t^2$ ($P<0.0001$) (channel 1); $%
0.256C(t)=588+0.0394t-0.000462t^2$ ($P<0.0001$) (channel 2); $%
0.256C(t)=394-0.0602t-0.000201t^2$ ($P<0.0001$) (channel 3); $%
0.256C(t)=205+0.0122t+0.000106t^2$ ($P<0.0001$) (channel 4). The fit to the
four channel light curves produces: $E_{0,C}=0.191keV$, $\alpha _{0,C}=0.398$%
, $\sigma _r=0.298$, $\sigma _d=0.181$, $t_1=2.98\times 10^5s$, $t_0=-10.1s$%
, and $C_0=1.90\times 10^4photon$. The reduced $\chi ^2$ of the fit is: $%
\chi _{com,\nu }^2=2.85$.

\textbf{GRB 920830:}

There are 7504 data points in total for GRB 920830 as well. In the same way
we divide this range of data uniformly into 1876 bins. The quadratic
polynomial fits to the background counts yield: $%
0.256C(t)=879-0.0493t-0.000242t^2$ ($P<0.0001$) (channel 1); $%
0.256C(t)=604-0.0233t-0.000118t^2$ ($P<0.0001$) (channel 2); $%
0.256C(t)=425-0.0407t+0.000108t^2$ ($P<0.0001$) (channel 3); $%
0.256C(t)=234-0.0568t+0.000175t^2$ ($P<0.0001$) (channel 4). The fit to the
four channel light curves produces: $E_{0,C}=0.220keV$, $\alpha _{0,C}=0.467$%
, $\sigma _r=0.489$, $\sigma _d=0.469$, $t_1=5.10\times 10^4s$, $t_0=-2.76s$%
, and $C_0=1.44\times 10^4photon$. The reduced $\chi ^2$ of the fit is: $%
\chi _{com,\nu }^2=3.10$.

\textbf{GRB 990816:}

The number of data points for GRB 990816 is 5856. We divide this range of
data uniformly into 1464 bins. The quadratic polynomial fits to the
background counts yield: $0.256C(t)=961-0.0529t-0.000160t^2$ ($P<0.0001$)
(channel 1); $0.256C(t)=689+0.0362t+0.0000328t^2$ ($P<0.0001$) (channel 2); $%
0.256C(t)=626+0.265t+0.000211t^2$ ($P<0.0001$) (channel 3); $%
0.256C(t)=426+0.354t+0.000150t^2$ ($P<0.0001$) (channel 4). The fit to the
four channel light curves produces: $E_{0,C}=0.166keV$, $\alpha _{0,C}=0.824$%
, $\sigma _r=0.0916$,$\sigma _d=0.111$, $t_1=2.95\times 10^5s$, $t_0=-3.03s$%
, and $C_0=3.13\times 10^5photon$. The reduced $\chi ^2$ of the fit is: $%
\chi _{com,\nu }^2=2.74$.

\begin{figure}[tbp]
\centering
\hspace*{1in}
\includegraphics[width=2.0in]{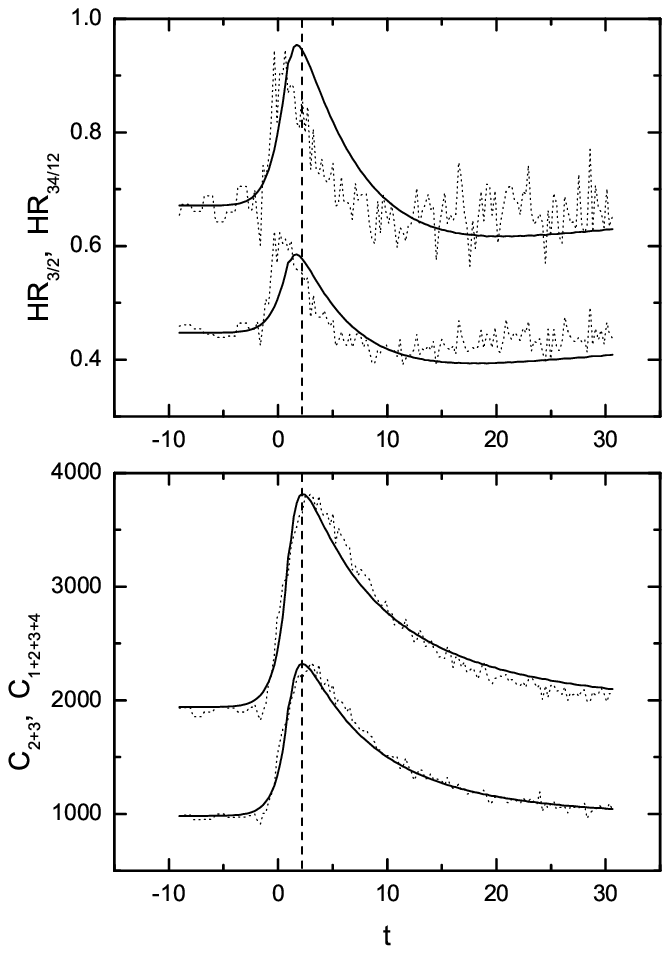}
\hspace*{1in}
\includegraphics[width=2.0in]{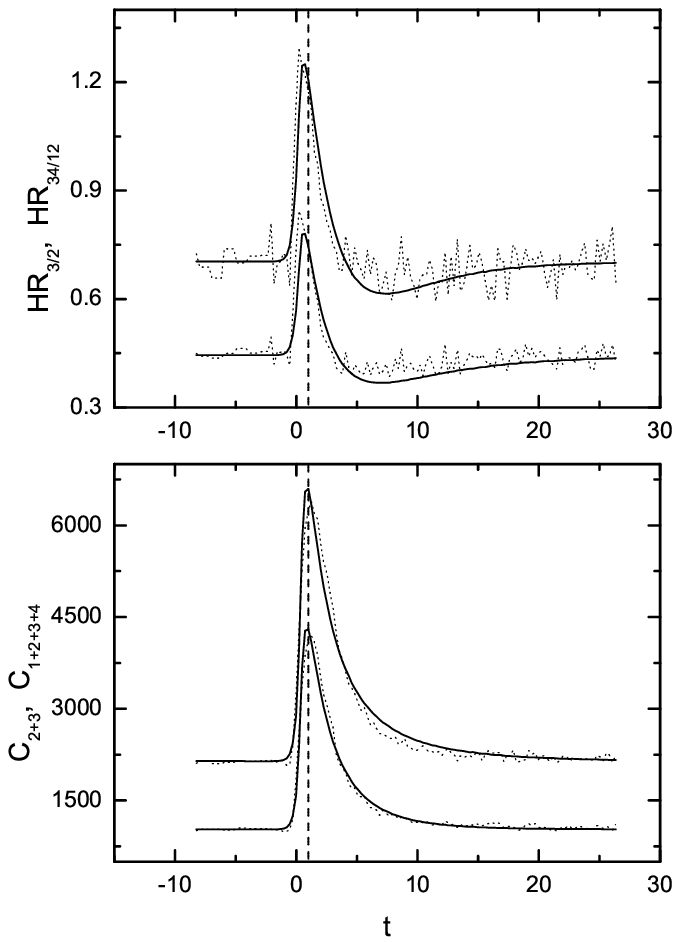}
\hspace*{1in}
\includegraphics[width=2.0in]{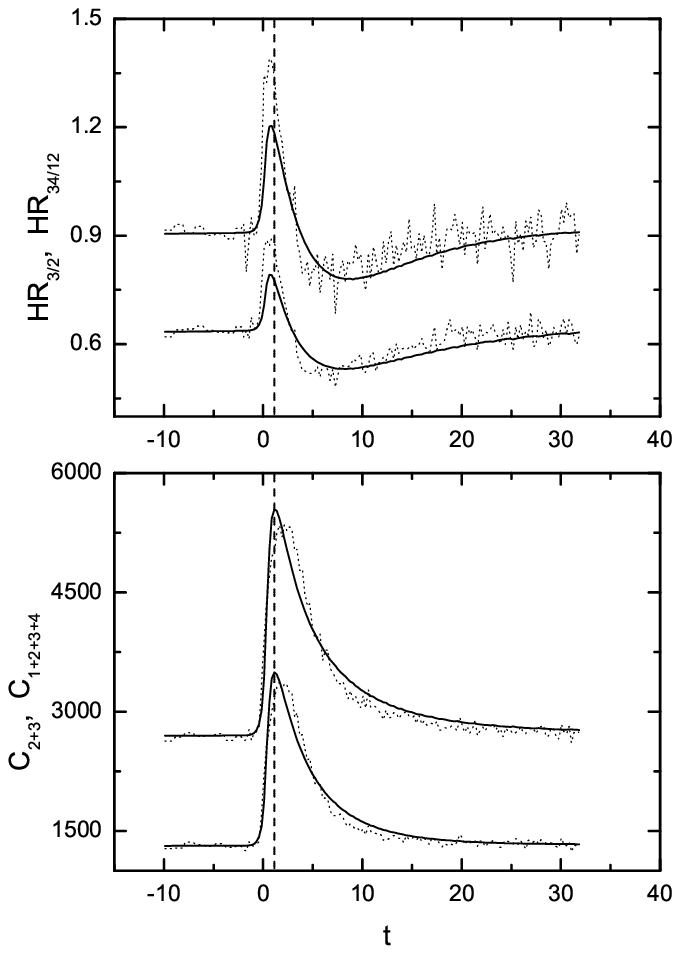}
\caption{Fits to the developments of the raw hardness ratios (upper panels)
and the multi-channel light curves (lower panels) of GRB 920216 (top left),
GRB 920830 (top right) and GRB 990816 (bottom) figure.
In the upper panels, solid
lines from the bottom to the top represent two expected hardness ratio
curves deduced from the four fitting light curves, which are defined by $%
HR_{34/12}\equiv (C_3+C_4)/(C_1+C_2)$ and $HR_{3/2}\equiv C_3/C_2$,
respectively, and dot lines from the bottom to the top stand for the
observed data of the corresponding hardness ratio curves respectively. In
the lower panels, solid lines from the bottom to the top represent two
expected multi-channel light curves deduced from the four fitting light
curves, which are defined by $C_{2+3}\equiv C_2+C_3$ and $C_{1234}\equiv
(C_1+C_2+C_3+C_4)$, respectively, and dot lines from the bottom to the top
stand for the observed data of the corresponding multi-channel light curves
respectively. The dash line in this figure denotes the position of the peak
of the expected curve of $C_{2+3}$.}
\end{figure}

As shown in Fig. 6, we present both the observed HRCs and the expected HRCs
which are deduced from the four fitting light curves, together with two
corresponding multi-channel light curves of the three sources.

One finds that the curvature effect could indeed produce the
characteristics of HRCs observed in the three bursts. Besides
this, we find that the observed hardness ratio tends to be harder
at the beginning of the pulses than what the curvature effect
could predict (which is represented by the corresponding fitting
curves) and softer at the late time of the pulses. This is called
a ``harder-leading'' problem. It indicates that if we believe that
the curvature effect is important as assumed above, then besides
the curvature effect there might be other effects being at work.
Nevertheless, as the figure shows, although it is possible that
there exist some other effects, {\it the curvature effect could
indeed be a key factor accounting for the characteristics the
observed HRCs of the three bursts show.}

\begin{figure}[tbp]
\centering
\hspace*{1in}
\includegraphics[width=3.0in]{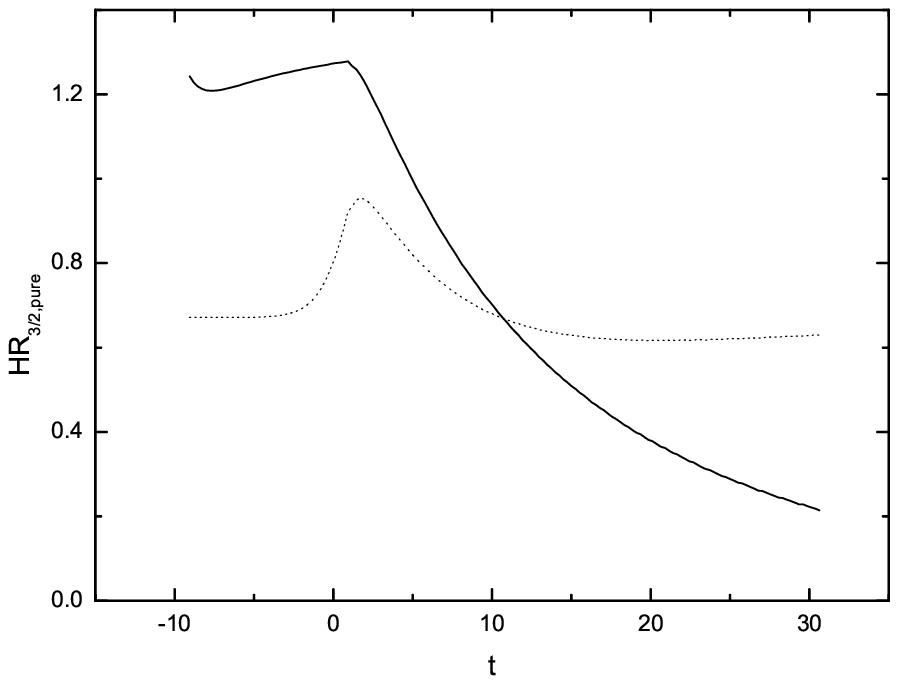}
\hspace*{1in}
\includegraphics[width=3.0in]{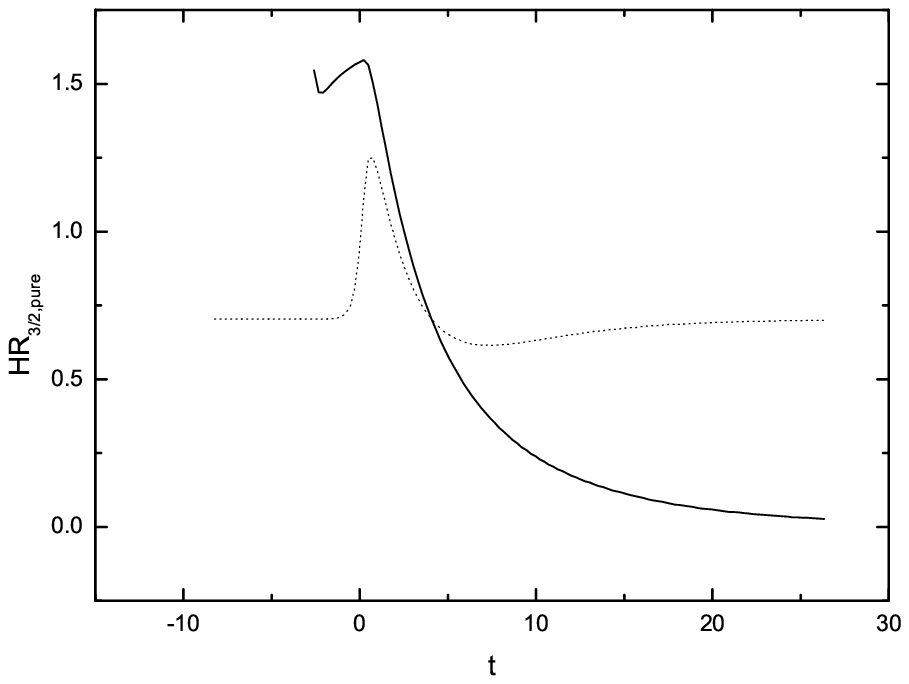}
\hspace*{1in}
\includegraphics[width=3.0in]{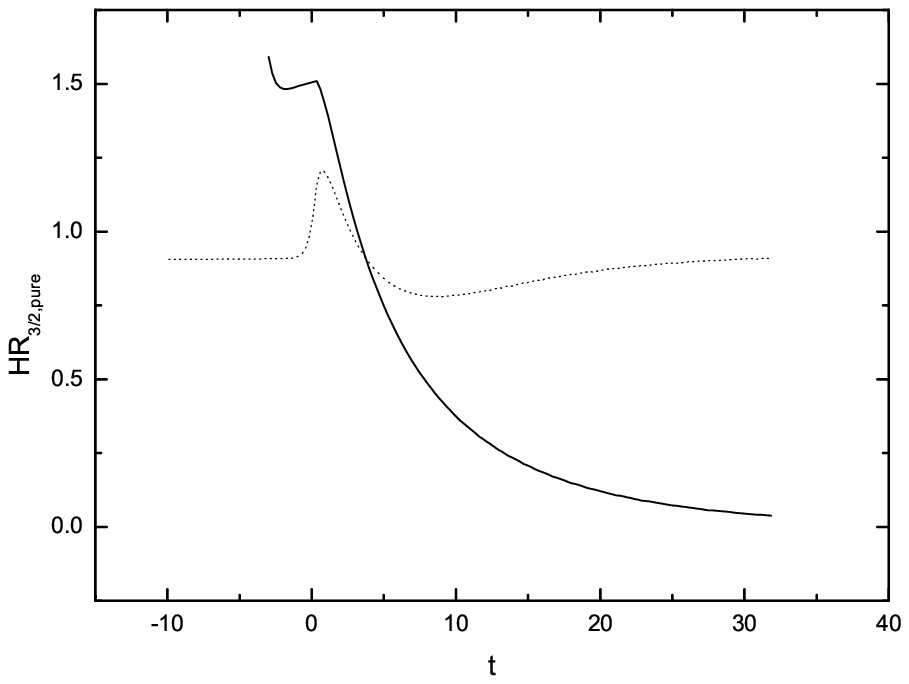}
\caption{Developments of the pure hardness ratios (the solid line) and raw
hardness ratio curves (the dotted line) for the three sources GRB 920216,
GRB 920830 and GRB 990816 (from top to bottom).}
\end{figure}


Pure hardness ratio curves of the three sources deduced from the four fitting
light curves of each burst are shown in Fig. 7. One finds that, in
the three cases, when the ``harder-leading'' problem is ignored the pure
hardness ratio curve would be hardest at the very beginning and then would
undergo a short period of drop-to-rise and later would decay continuously
until the pulse dies away, as predicted above. The short period of
drop-to-rise is a characteristic of the curvature effect. It is unclear if
this characteristic still holds when the mechanism leading to the
``harder-leading'' problem is at work. However, we believe that this
characteristic could be expected in cases when the ``harder-leading''
problem is very insignificant. We wonder if one could check this with some
other sources where the scattering of data is small enough so that the pure
hardness ratio could be well measured at the very beginning of the pulses.

As pointed out by [17], the phenomenon of the softening
of the spectra with time revealed by statistical analysis is quite general
which is independent of any changes in the intrinsic spectrum and therefore
independent of the physical environment where the pulses are produced. The
observed phenomenon is generally in agreement with the evolutionary curve of
the pure hardness ratio illustrated in Fig. 7. But in terms of the raw
hardness ratio, we find a soft-to-hard-to-soft feature instead. This is a
characteristic associated with the drop-to-rise-to-drop feature shown in the
evolutionary curve of the pure hardness ratio (see Fig. 7), and therefore
the two features are alternatives which could serve as an indicator of the
curvature effect (note that they are not independent).

\section{Discussion and conclusions}

The first question put forward is that: are the three bursts isolated cases
showing the characteristics predicted by the curvature effect? This could be
easily answered when one examines the HRC for more bursts. Listed in the
tables of [26] are seven other FRED pulse bursts, GRB910721
(\#563), GRB920925 (\#1956), GRB930214c (\#2193), GRB930612 (\#2387),
GRB941026 (\#3257), GRB951019 (\#3875) and GRB951102B (\#3892), which were
also intensely studied elsewhere [18,21]. Let us
show the HRC curve for these bursts and examine their characteristics.

\begin{figure}[tbp]
\centering
\includegraphics[width=2.0in]{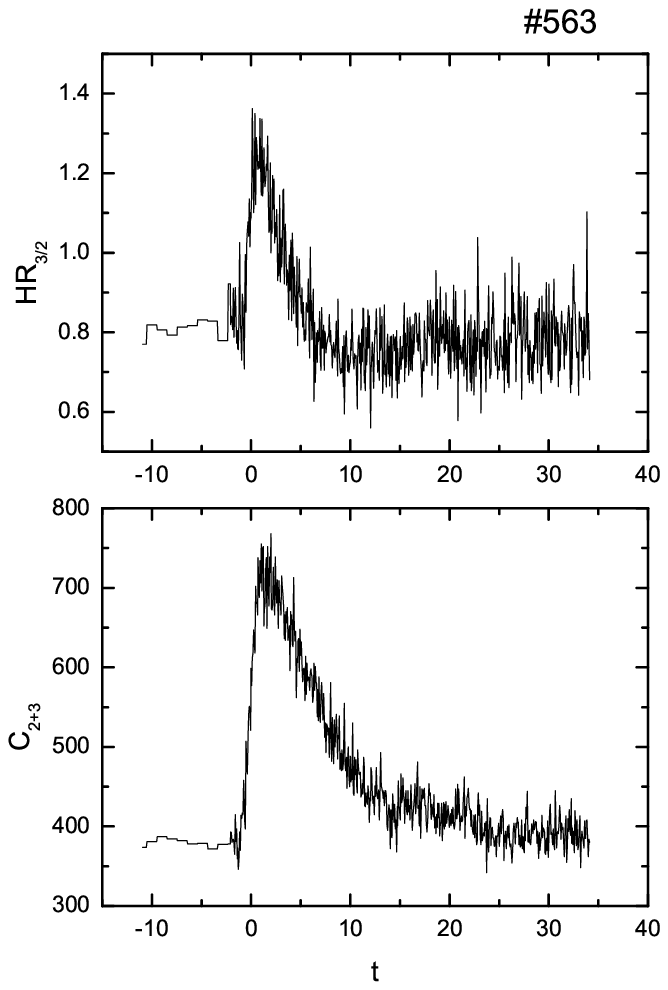} \includegraphics[width=2.0in]{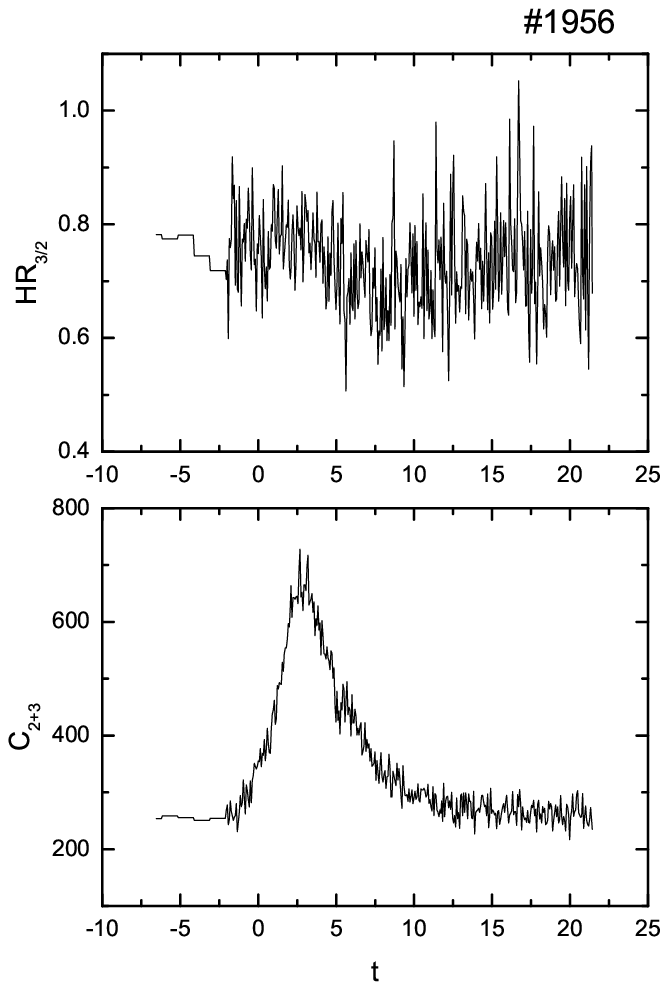}
\includegraphics[width=2.0in]{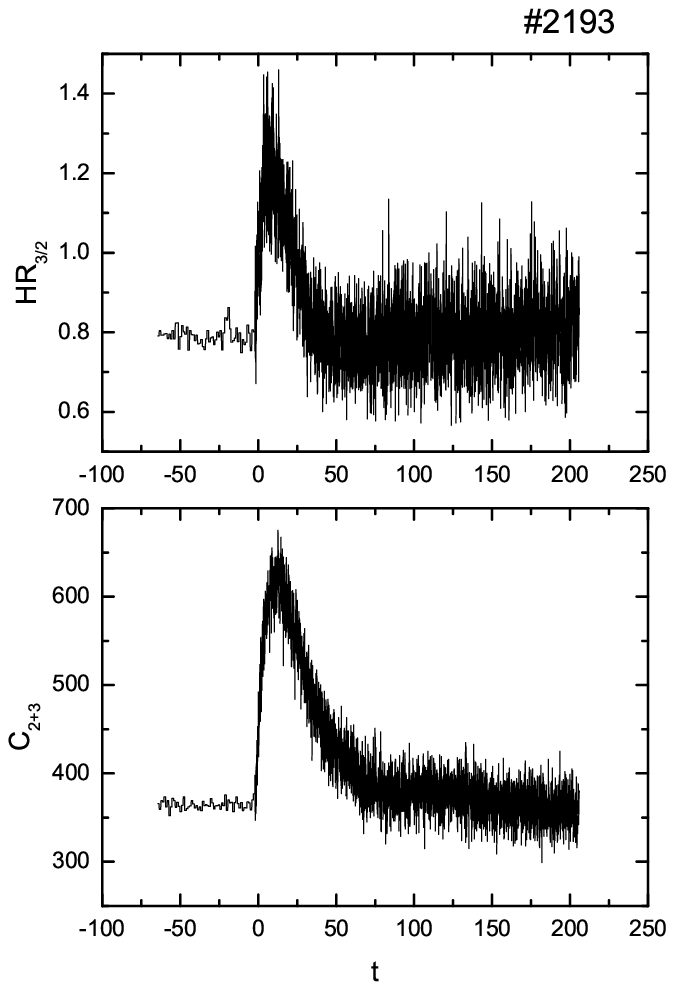} \includegraphics[width=2.0in]{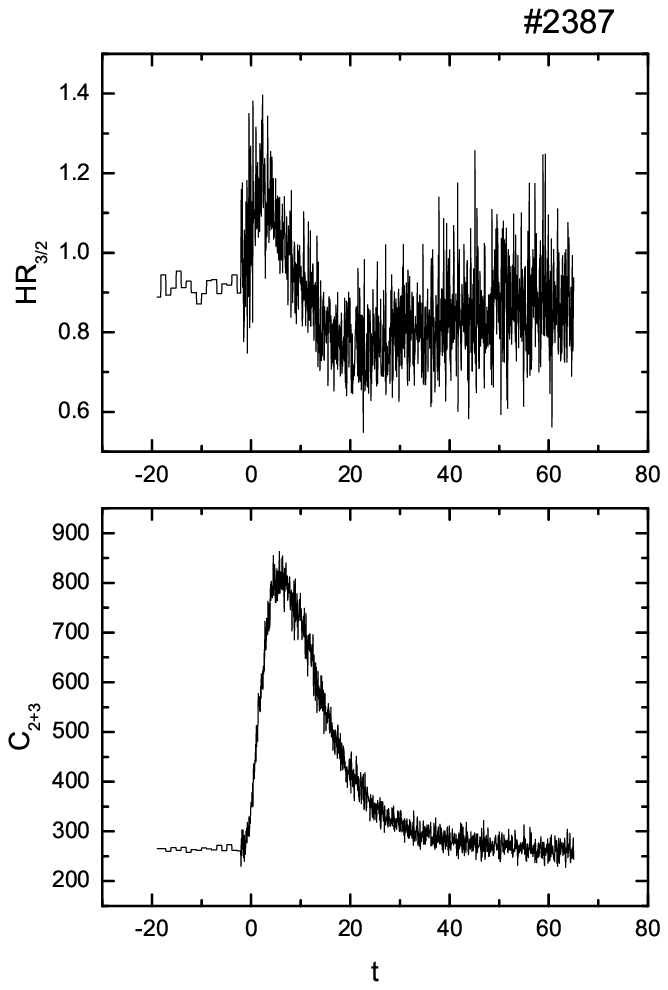}
\includegraphics[width=2.0in]{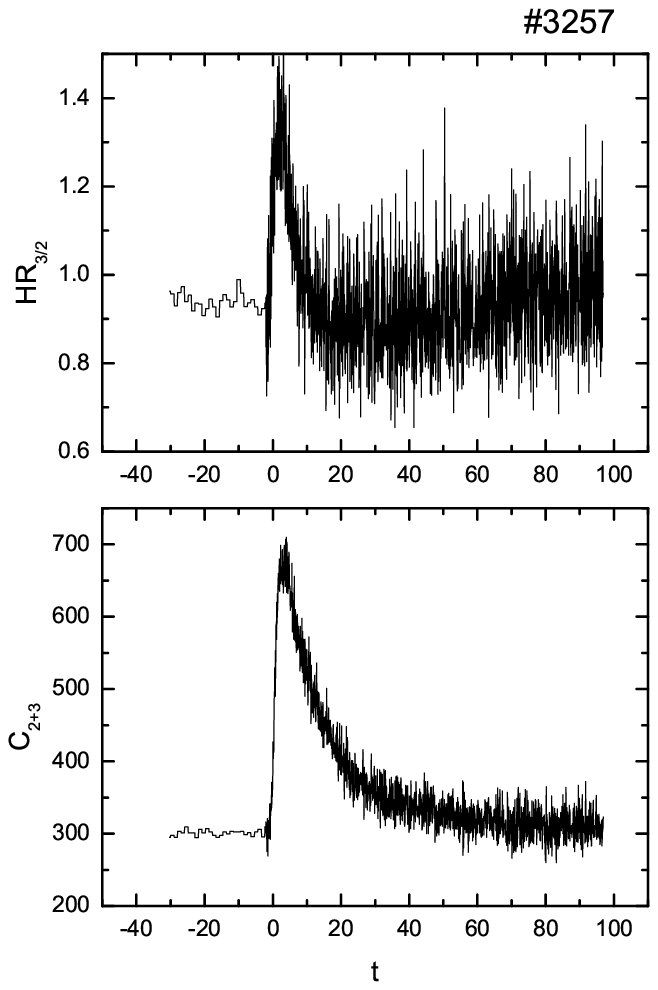} \includegraphics[width=2.0in]{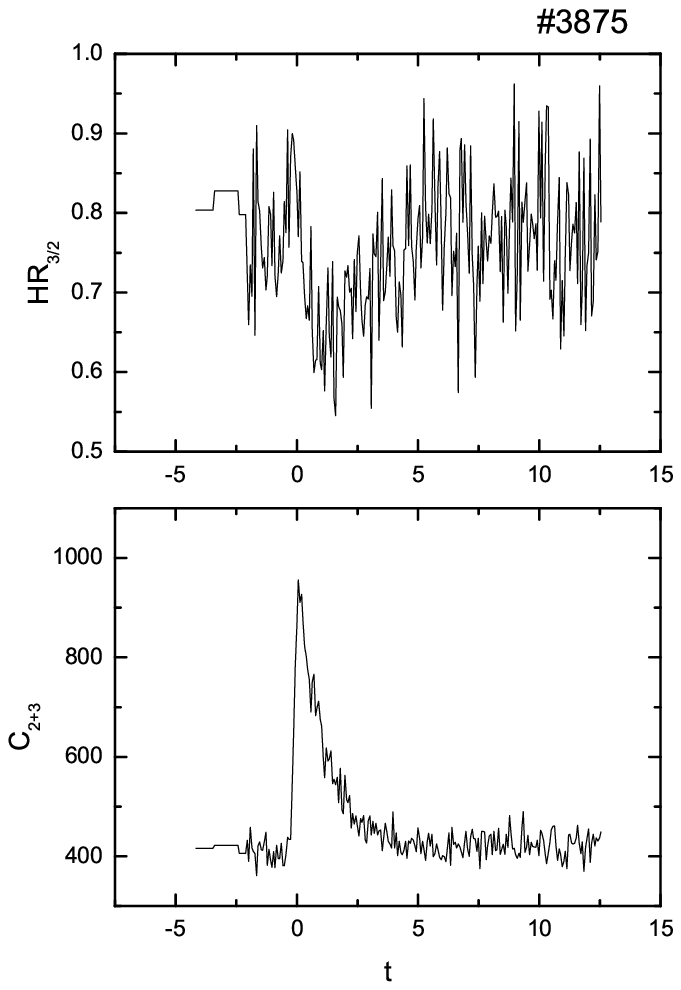}
\includegraphics[width=2.0in]{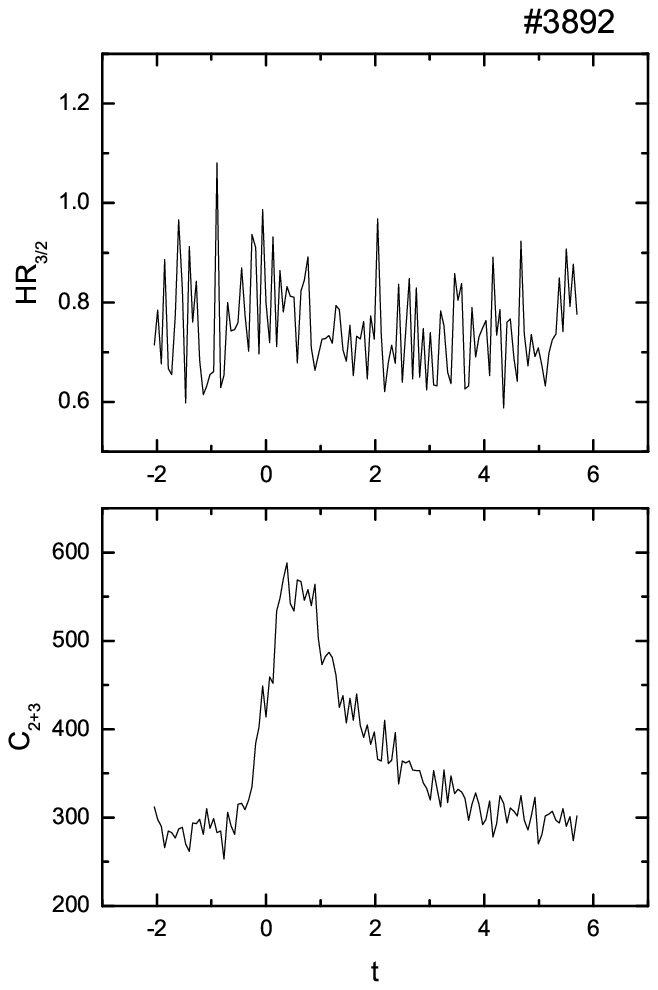}
\caption{Developments of the raw hardness ratios and the bi-channel light
curves which includes count rates of both the second and third channels of
the seven bursts.}
\end{figure}

Fig. 8 shows the HRC as well as the corresponding bi-channel light
curve of the seven bursts. We find that, for GRB910721,
GRB930214c, GRB930612, GRB941026, the characteristics predicted by
the typical hard burst are observed (see Fig. 1, the 4th and 5th
panels counting from the top). For GRB920925 and GRB951019, a
possible sinkage feature is observed, although the fluctuation of
background is very large. Compared with the theoretical
predictions, these two bursts might be relatively soft (see Fig.
2). This conjecture could be checked if one is able to remove or
largely ease the great fluctuation observed in their HRCs (by the
smooth of data or other approaches; the issue is beyond the scope
of this paper). For GRB951102B, no signals are detected due to the
great chaos and therefore we cannot tell if the curvature effect
is or is not at work. We suspect that this might not only be due
to fluctuation, but also be due to the small signal it might
possibly possess in its HRC.

The second question concerns that what causes the so-called
``harder-leading'' problem. It should be reminded that several other
possible effects are ignored in the above analysis. Among them there is an
economic one accounting for this effect, which is the development of the
rest frame spectrum when it is harder, itself, at the beginning and softer
at the late time of the pulses. The mechanism leading to this seems
expectable. In the case of a Comptonized radiation, electrons of the inner
shell that scattering photons in the outer shell would possess a larger
value of speed relative to the expanding fireball surface at the early time
of shocks and a smaller value later. In the case of a synchrotron radiation,
electrons that radiate are expected to gain larger acceleration at the
beginning of a shock than at later times. Obviously, how the rest frame
spectrum evolves depends on what mechanisms dominate in the corresponding
period.

In this paper, we use a Comptonized spectrum instead of the Band
function to perform the fit due to several reasons. The first is
that the Comptonized spectrum could indeed represent some GRB
spectra [39]. The second is that we would like to know if a real
mechanism could be employed to describe (even roughly) the
observed HRC. The third is that there are only two free parameters
in the Comptonized spectrum, while in the Band function form there
are three. From Fig. 6 one finds that, since the fit is quite
satisfied, {\it ignoring the possible effect arising from the
deviation of the Comptonized spectrum from the real rest frame one
may not give rise to a severe problem.} And it seems unlikely that
this effect could lead to the ``harder-leading'' problem so long
as there is a more economic mechanism accounting for the latter.

It is noticed that, the peak energy alone cannot determine the hardness
ratio (to determine the hardness ratio, one needs to know the peak energy as
well as the low and high energy indexes). Any of the two quantities could
not replace the other. The development of the hardness ratio reflects only
one aspect of the spectral evolution, while that of the spectral peak energy
would reflect another aspect. Thus, to understand the development of the
spectrum in a more detail, it is necessary to study the evolutionary curve
of the peak energy as well. We expect that such an investigation could
independently test the curvature effect.

There has been a consensus that due to the relativistic aberration
of light, isotropically emitted radiation in the co-moving frame
will be beamed into a cone with opening angle $\theta \sim
1/\Gamma $ and therefore only photons emitted from the fireball
surface within this cone will be detectable by the observer [40].
However, as shown in [18], emission from a cone of opening angle
$\theta =1/\Gamma $ would lead to a FRED pulse light curve with a
cutoff tail (when the local pulse is extremely narrow) or a
turnover feature (when the local pulse is not so narrow). It
suggests that contribution of the emission from the area of
$\theta
>1/\Gamma $ to the tail of the light curve is important {\it (while it is indeed
negligible in the main part of the light curve}; see Figs. 1 and 2
in [18]). This effect might have an impact on the HRC. In this
paper, we take $\theta =0.03$ according to [38], which is much
larger than $1/\Gamma =0.005$ for $\Gamma =200$ adopted in this
paper. Since the effects associated with $\theta =0.03$ and
$\theta =\pi /2$ are not distinguishable, the conclusions obtained
in this paper hold for both cases of a spherical fireball and an
uniform jet when its opening angle is large enough.

The plan is that when a more flexible form of local pulses is
adopted, one will be able to obtain a better fit. However, the
evidence of curvature effect presented above is obvious enough.
{\it The fact that adopting local pulse (2) can simply account for
the observed data of the three bursts when merely taking into
account the curvature effect shows explicitly that the effect
could indeed serve as the main cause of the observed HRC
characteristics of some bursts, especially for those with FRED
pulses.}

It is obvious that an intrinsic hard-to-soft spectral
evolution could account for the phenomenon of the spectral
softening with time which were generally observed in GRBs. Thus,
the above analysis suggests only that the curvature effect is a
plausible origin of the phenomenon. To tell if a burst suffers
effects other than the curvature one, one needs to investigate if
any of these effects could give rise to the characteristics the
light curve and the HRC of the burst possesses. This will take
some time and it deserves.

Based on the above analysis and discussion, the main
conclusions of this paper are as follows. Due to the curvature
effect, the evolutionary curve of the pure hardness ratio would
peak at the very beginning of the curve, and then would undergo a
drop-to-rise-to-decay phase. In the case of the raw hardness
ratio, the curvature effect would give rise to several types of
evolutionary curve, depending on how hard is a burst. For the
typical soft burst with $\Gamma =20$ (when assuming a rest frame
Comptonized radiation with indexes of $\alpha _{0,C}=-0.6$ and
$\nu _{0,C }=0.55keVh^{-1}$), an upside-down pulse would be
observed; for the typical hard burst with $\Gamma =200$, a
pulse-like profile with a sinkage in its decaying phase would be
observed; for the typical very hard burst with $\Gamma =2000$, a
pulse-like profile without a sinkage in its decaying phase would
be observed. In the case that the raw hardness ratio evolutionary
curve is a pulse-like one, as shown in the typical hard and very
hard bursts, its peak would appear in advance of that of the
corresponding light curve. The features of raw HRCs predicted in
the case of the typical hard burst are observed in GRB 920216, GRB
920830, and GRB 990816. {\it A fit to these bursts shows that the
curvature effect alone could indeed account for the
characteristics of the HRCs observed in some bursts.} 

In addition, we find that the observed hardness ratio tends to be
harder at the beginning of the pulses than what the curvature
effect could predict and softer at the late time of the pulses.
According to the discussion, we tend to believe that this is
an evidence showing the existence of intrinsic hard-to-soft
radiation which probably be due to the
acceleration-to-deceleration mode of shocks. Illustration of the
corresponding curves of seven other bursts shows that the
characteristics of HRCs observed in the three bursts are common 
(four of seven cases) in FRED pulse sources and the features predicted by
the typical soft burst are seen in two cases.

\begin{acknowledgments}
This work was supported by the National Science Fund (10125313), the 
National Natural Science Foundation of China (No. 10573005 and  No. 10273019), 
and by a Guangdong Provience fund (Q02114). We also thank the Guangzhou Education 
Bureau and Guangzhou Science and Technology Bureau for financial support.
\end{acknowledgments}

\bibliography{}

[1]{} J. Goodman, Astrophys. J. {\bf 308}, L47 (1986)

[2]{} B. Paczynski, Astrophys. J. {\bf 308}, L43 (1986)

[3]{} J. H. Krolik and E. A. Pier, Astrophys. J. {\bf 373}, 277 (1991)

[4]{} P. Meszaros and M. J. Rees, Astrophys. J. {\bf 502}, L105 (1998)

[5]{} C. J. Hailey, F. A. Harrison and K. Mori, Astrophys. J. {\bf 520}, L25 (1999)

[6]{} Y.-P. Qin, Astron. Astrophys. {\bf 396}, 705 (2002)

[7]{} Y.-P. Qin, Astron. Astrophys. {\bf 407}, 393 (2003)

[8]{} M. Tavani, Astrophys. J. {\bf 466}, 768 (1996)

[9]{} N. Lloyd and V. Petrosian, Astrophys. J. {\bf 543}, 722 (2000)

[10]{} A. Panaitescu and P. Meszaros, Astrophys. J. {\bf 544}, L17 (2000)

[11]{} E. P. Liang, Astrophys. J. {\bf 491}, L15 (1997)

[12]{} D. Band et al., Astrophys. J. {\bf 413}, 281 (1993)

[13]{} G. J. Fishman et al. Astrophys. J. Suppl. Ser. {\bf  92}, 229 (1994)

[14]{} J. P. Norris et al., Astrophys. J. {\bf 459}, 393 (1996)

[15]{} D. Kocevski, F. Ryde and E. Liang, Astrophys. J. {\bf 596}, 389 (2003)

[16]{} E. E. Fenimore, C. D. Madras and S. Nayakshin, Astrophys. J. {\bf 473}, 998 (1996)

[17]{} F. Ryde and V. Petrosian, Astrophys. J. {\bf 578}, 290 (2002)

[18]{} Y.-P. Qin et al., Astrophys. J. {\bf 617}, 439 (2004)

[19]{} J. P. Norris et al., Astrophys. J. {\bf 534}, 248 (2000)

[20]{} E. E. Fenimore et al., Astrophys. J. {\bf 448}, L101 (1995)

[21]{} R. J. Nemiroff, Astrophys. J. {\bf 544}, 805 (2000)

[22]{} Y.-P. Qin and R.-J. Lu, Mon. Not. R. Astron. Soc. {\bf 362}, 1085 (2005)

[23]{} Y.-P. Qin et al., Astrophys. J. {\bf 632}, 1008 (2005)

[24]{} R.-F. Shen, L.-M. Song and Z. Li, Mon. Not. R. Astron. Soc. {\bf 362}, 59 (2005)

[25]{} R.-J. Lu et al., Mon. Not. R. Astron. Soc. {\bf 367}, 275 (2006)

[26]{} L.-W. Jia and Y.-P. Qin, Mon. Not. R. Astron. Soc. {\bf 631}, L25 (2005)

[27]{} Z.-Y. Peng et al., Mon. Not. R. Astron. Soc. {\bf 368}, 1351 (2006)

[28]{} E. W. Liang et al., Astrophys. J. {\bf 646}, 351, (2006)

[29]{} J. P. Norreis et al., Astrophys. J. {\bf 301}, 213 (1986)

[30]{} V. E. Kargatis et al. Astrophys. Space Sci. {\bf 231}, 177 (1995)

[31]{} F. Ryde and R. Svensson, Astrophys. J. {\bf 529}, L13 (2000)

[32]{} F. Ryde and R. Svensson, Astrophys. J. {\bf 566}, 210 (2002)

[33]{} L. Borgonovo and F. Ryde, Astrophys. J. {\bf 548}, 770 (2001)

[34]{} P. N. Bhat et al., Astrophys. J. {\bf 426}, 604 (1994)

[35]{} For full information of the data of the burst and transient source
experiment, see the web site of NASA's CGRO Science Support Center, \\
http://cossc.gsfc.nasa.gov/docs/cgro/batse/.

[36]{} Concatenated 64-ms burst data in ASCII format of the burst and transient
source experiment are available at \\
http://cossc.gsfc.nasa.gov/docs/cgro/batse/batseburst/sixtyfour$\_$ms/index.html.

[37]{} I. Mitrofanov et al., in Gamma-ray bursts -- {\it Observations,
Analyses and Theories}, edited by C. Ho, R. I. Epstein, and E. E. Fenimore
(Cambridge University Press, Cambridge, England, 1992), p. 209.

[38]{} A. S. Friedman and J. S. Bloom, Astrophys. J. {\bf 627}, 1 (2005)

[39]{} R. D. Preece et al., Astrophys. J. Suppl. Ser. {\bf 126}, 19 (2000)

[40]{} F Ryde et al., Astron. Astrphys. {\bf 411}, L331 (2003).

\end{document}